# Hikami-Larkin-Nagaoka (HLN) Fitting of Magneto Transport of $Bi_2Se_3$ Single Crystal in Different Magnetic Field Ranges

Deepak Sharma[1,2,a], P. Rani[1], P.K. Maheshwari[1,2], V. Nagpal[3], R.S. Meena[1,2], S.S. Islam[4], S. Patnaik[3] and V.P.S. Awana[1,2]

[1]*CSIR-National Physical Laboratory, K.S. Krishnan Marg, New Delhi-110012, India*
[2]*Academy of Scientific and Innovative Research (AcSIR), Ghaziabad, 201002, India*
[3]*School of Physical Sciences, Jawaharlal Nehru University, New Delhi-110067, India*
[4]*Centre of nanoscience and nanotechnology, Jamia Millia Islamia, New Delhi-110025, India*

[a] Corresponding author: sharmadeepak1111@gmail.com

**Abstract.** We report the detailed study of structural/micro-structuraland high magnetic field magneto transport propertiesof $Bi_2Se_3$single crystal. $Bi_2Se_3$singlecrystal is grown through conventional solid-state reaction route via the self-flux method. Rietveld analysis on Powder X-ray Diffraction (PXRD)showed that the studied $Bi_2Se_3$ crystal is crystallized in single-phasewithout any impurity. The surface morphology analyzed through Scanning Electron Microscopy (SEM) study which shows that as-grown single crystal exhibit layered type structure and the quantitative weight% of the atomic constituents (Bi and Se) are found to be closeto the stoichiometric amount in energy-dispersive X-ray spectroscopy (EDS) analysis. Low temperature (2.5K) magneto-resistance (MR) exhibited a v-type cusp around origin at lower magnetic field, which is the sign of weak anti-localization(WAL) effect. Further, $Bi_2Se_3$ single crystal magneto conductivity data is fitted by well-known HLN equation in different magnetic field range of 2Tesla, 4Tesla and 6Tesla and the resultant found that the conduction mechanism of $Bi_2Se_3$ is dominated by WAL state.

## INTRODUCTION

Topological Insulator (TI) is the new state of quantum matter having protected conducting surface states; on the other hand, the bulk is the insulator. These surface states are Time Reversal Symmetry (TRS) protected and do have strong Spin-Orbit Coupling (SOC). $Bi_2Se_3$ is the binary tetradymite compounds in which the strong SOC is present and is having single Dirac cone in their energy versus momentum diagram at Γ point [1-3]. $Bi_2se_3$ has attracted lot of attention due to their various quantum phenomenonas Weak Anti Localization(WAL), Aharonov-Bohm (AB) oscillation and high field linear Magneto Resistance (MR) associated with their conducting surface states [4, 5].WAL is considered due to the suppression of backscattering of carriers by the π-Berry phase of topological surface states and helical spin momentum locking as reported earlier [4, 5].

In the present short article, we report the successful growth and structural/micro-structural details and high field (10 Tesla) magneto conductivity analysisby HLN fitting of $Bi_2Se_3$single crystal.

## EXPERIMENTAL DETAILS

High-quality $Bi_2Se_3$ single crystal has been synthesized by the self-flux method through the conventional solid-state reaction route [6]. High-purity (99.99%) bismuth (Bi) and selenium (Se) were weighed accurately in their stoichiometric ratio, well mixed and ground thoroughly inside a glove box under high-purity Ar (Argon) atmosphere to avoid oxidation of the any element. The homogeneously mixed powder then waspressed in form of a rectangular pellet using a hydraulic press under a pressure of 50kg/cm$^2$ and then vacuum-sealed($10^{-5}$Torr) into high quality quartz tube. The sealed tube was then sintered inside a tube furnace with a rate of 2$^o$C/min. upto 950$^o$C, kept there for 24 h and then slowly cooled down to 650°C at a rate of 2°C/h. Further, the tube furnace was allowed to cool to room temperature naturally. The obtained sample was shiny and silver in color, which was mechanically cleaved for further characterizations.

XRD pattern had been performed using a Rigaku Made Mini Flex II X-ray diffractometer and SEM study followed by EDS were performed on Bruker made scanning electron microscope. Further, magneto resistance measurements were carried out by a conventional four-probe method on a Physical Property Measurement System (PPMS-10Tesla) using a close cycle refrigerator.

## RESULT AND DISCUSSION

Figure 1(a) shows the Rietveld fitted room-temperature XRD pattern of a crushed piece of $Bi_2Se_3$ crystal. All peaks are well fitted confirms that the sample is crystallized in single phase having arhombohedral structure within the R̄3m space group [7]. The lattice parameters as obtained from the Rietveld refinement are a = b = 4.15609(3) Å, and c = 28.73611(8) Å andthe values of α, β and γ are 90°, 90°, and 120° respectively.

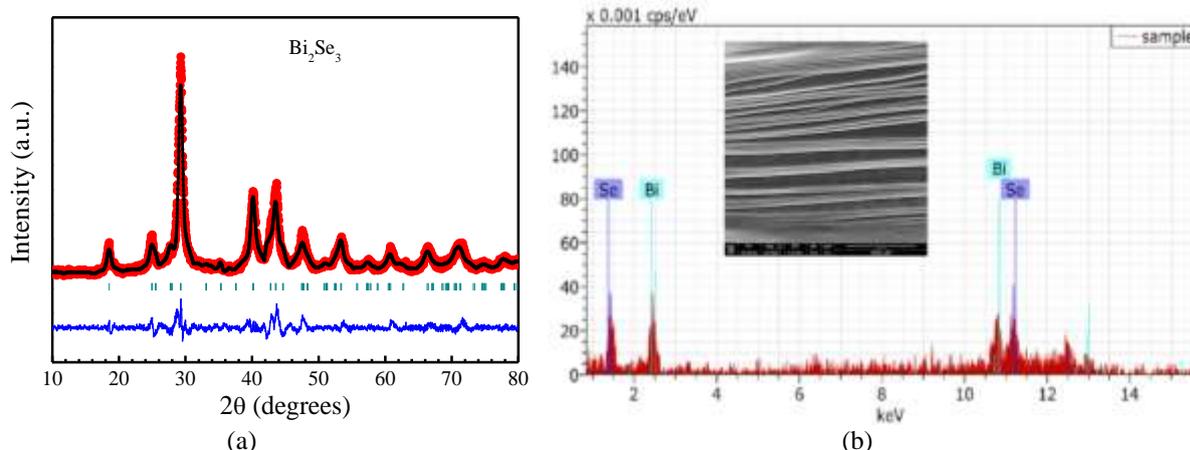

**FIGURE 1.** (a) Rietveld fitted X-ray diffraction pattern and (b) EDS analysis of $Bi_2Se_3$ single crystal and inset is SEM image of same.

SEM and EDS analyze the surface morphological characteristics, as well as the chemical composition of the as grown $Bi_2Se_3$ single crystal. Fig. 1(b) exhibits the EDS spectrum with SEM image in inset view of as grown $Bi_2Se_3$ single crystal. The SEM image of the studied singlecrystal exhibits layered type crystal morphology [8], and the quantitative weight% of the atomic constituents (Bi and Se) is found to be near the stoichiometric amount. Apparently, the studied crystal seems to be pure, composed of only Bi and Se atomic constituents present in the studied sample.

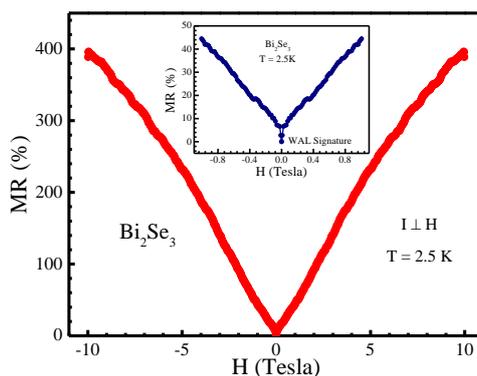

**FIGURE 2.** Magnetoresistance (MR%) as a function of varying applied magnetic field from -10Tesla to +10Tesla for pure $Bi_2Se_3$ single crystal at 2.5K temperature, inset shows the zoomed part of same in low field range from -1Tesla to +1Tesla.

Figure 2 shows the magneto resistance (MR%) as a function of varying applied magnetic fields from +10Tesla to -10Tesla for $Bi_2Se_3$ single crystal at 2.5K temperature. Almost a linear curve is found with increase in magnetic field. A clear v-type cusp around origin is seen at lower magnetic field. Thus, at lower magnetic field a sharp dip like positive MR is observed in absence of any magnetic scattering. This type of behavior is known for the signature of WAL effect in earlier reports[9, 10]. The inset of fig. 2 shows the closed view of MR% upto ±1 Tesla applied magnetic field.

The surface states dominated conductivity of topological insulators is known to follow the HLN (Hikami-Larkin-Nagaoka) equation as below [11]

$$\Delta\sigma(H) = \sigma(H) - \sigma(0) = -\frac{\propto e^2}{\pi h}\left[\ln\left(\frac{B_\varphi}{H}\right) - \Psi\left(\frac{1}{2} + \frac{B_\varphi}{H}\right)\right]$$

Here, $\Delta\sigma(H)$ represents the change of magneto-conductivity, $\sigma(0)$ conductivity at zero magnetic field, **α** is a coefficient signifying the type of localization, **e** denotes the electronic charge, **h** represents the Planck's constant, **Ψ** is the digamma function, **H** is the applied magnetic field, $B_\varphi = \frac{h}{8e\pi(l_\varphi^2)}$ is the characteristic magnetic field and $l_\varphi$ is the phase coherence length. The HLN fitted magneto-conductivity plots for $Bi_2Se_3$ single crystal at 2.5K in various field ranges of 2, 4 and 6 Tesla are shown in figure 3. The obtained values of α are -0.48, -0.45 and -0.41 at 2, 4 and 6 Tesla respectively. On the other hand the values of $l_\phi$ are 81.58, 88.42 and 99.44 nm at 2, 4 and 6 Tesla respectively. The value of α is associated with different types of localizations (WL, WAL or both WL and WAL). The overall value of α decides the type of spin-orbit interactions (SOI and magnetic scattering) [11]. In present case the value of α is close to -0.50, which is known to be a strong SOI case without any magnetic scattering [11]. As it is near to -0.50 indicates WAL contribution of $Bi_2Se_3$ crystal in conduction mechanism, which is in agreement with MR(%) measurement (shown in fig. 2). The HLN fitted values of α and $l_\phi$ are tabulated in Table 1. There is general trend that the value of pre-factor α increases with increase in fitting ranges of the field. As well, the values of phase coherence length ($l_\phi$) increase with increase in fitting ranges of the magnetic field.

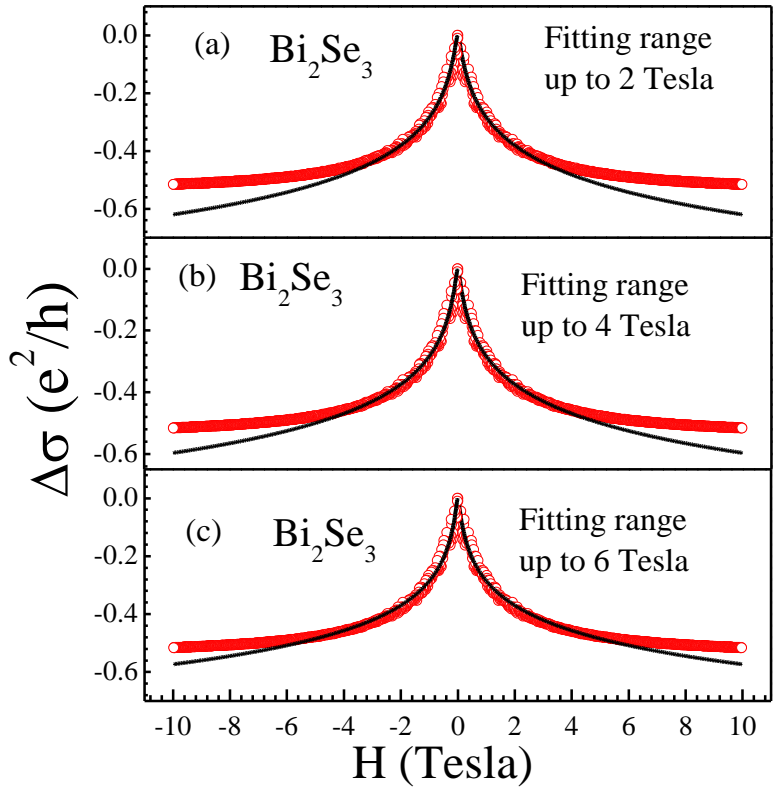

**FIGURE 3**. HLN fitting in the field range of (a) up to 2Tesla (b) up to 4Tesla (c) up to 6Tesla for $Bi_2Se_3$ crystal

**TABLE 1:** HLN fitted values of the pre-factor (α), phase coherence length ($l_\phi$) and fitted parameter ($R^2$) at 2.5K temperature for $Bi_2Se_3$ single crystal in different magnetic field range.

| Magnetic Field Range | pre-factor (α) | Phase coherence length ($l_\phi$) (nm) | Fitted parameter ($R^2$) |
|---|---|---|---|
| ±2 Tesla | -0.48 | 81.58 | 0.9691 |
| ±4 Tesla | -0.45 | 88.42 | 0.9855 |
| ±6 Tesla | -0.41 | 99.44 | 0.9849 |

## CONCLUSION

The $Bi_2Se_3$ crystal has been grown successfully through the conventional solid-state reaction route. Rietveld fitting of PXRD confirms the phase purity of asgrown crystal. The obtained results from SEM and EDS measurement confirm the growth of single crystal, which exhibits layered structure and the quantitative weight% of the atomic constituents (Bi and Se) are close to the stoichiometric amount. Magneto-transport measurement is carried out up to ±10 Tesla, a v-type cusp around origin is seen at lower magnetic field, which shows the signature of WAL effect, in the $Bi_2Se_3$ sample at temperature around 2.5 K. The change in magneto-conductivity has been estimated with the HLN fitting, the values of α (type of localization) and $l_\phi$ (Phase coherence length) obtained which validate the WAL state in conduction mechanism of $Bi_2Se_3$ crystal.

## ACKNOWLEDGMENT

The authors would like to thank Director NPL, India for his keen interest in the present work. Author would like to thanks CSIR, India for research fellowship and AcSIR-NPL for Ph.D. registration.

## REFERENCES

1. H. Zhang, C. X. Liu, X. L. Qi, X. Dai, Z. Fang and S. C. Zhang, Nat. Phys. **5**, 438-442 (2009).
2. Y. Ando and L. Fu, Annual Rev. Cond. Matt. Phys. **6**, 361-388 (2015).
3. M. Z. Hasan and C. L. Kane, Rev. Mod. Phys. **82**, 3045-3067 (2010).
4. H. Z. Lu, S. Q. Sen, Phys. Rev. B **84**, 125138-(1-8) (2011).
5. J.G. Checkelsky, Y. S. Hor, M. H. Liu, D. X. Qu, R.J. Cava, N.P. Ong, Phys. Rev. Lett. **103**, 246601-(1-4) (2009).
6. R. Sultana, P. Neha, R. Goyal, S. Patnaik and V.P.S. Awana, J. Magn. Mag. Matt. **428**, 213-218 (2017).
7. R. Sultana, P.K. Maheshwari, B. Tiwari and V.P. S. Awana, Mater. Res. Express **5**, 016102 (2018).
8. R. Sultana, G. Gurjar, S. Patnaik and V.P.S. Awana, Mater. Res. Express **5**, 046107 (2018).
9. S.X. Zhang, R.D. McDonald, A. Shekhter, Z.X. Bi, Y. Li, Q.X. Jia and S.T. Picraux, Appl. Phys. Lett. **101**, 202403-(1-4) (2012).
10. H. T. He, G. Wang, T. Zhang, I.K. Sou, G.K. L. Wong and J.N. Wang, Phy. Rev. Lett. **106**, 166805-(1-4) (2011).
11. S. Hikami, A. Larkin, Y. Nagaoka, Prog. Theor. Phys. **63**, 707-710 (1980).